\documentclass[letterpaper,english,reprint, aps]{revtex4-2}
\usepackage[T1]{fontenc}
\usepackage[utf8]{inputenc}
\usepackage{babel}
\usepackage{verbatim}
\usepackage{float}
\usepackage{amsmath}
\usepackage{amssymb}
\usepackage{graphicx}
\usepackage{booktabs}
\usepackage{graphicx}
\usepackage{gensymb}
\usepackage{subscript}
\usepackage[unicode=true,pdfusetitle,
bookmarks=true,bookmarksnumbered=false,bookmarksopen=false,
breaklinks=false,pdfborder={0 0 1},backref=false,colorlinks=false]
{hyperref}
\usepackage{xcolor}

\makeatletter

\newcommand{\lyxmathsym}[1]{\ifmmode\begingroup\def\b@ld{bold}
	\text{\ifx\math@version\b@ld\bfseries\fi#1}\endgroup\else#1\fi}

\usepackage{geometry} 
\usepackage{amsmath}
\usepackage{xparse}

\newmuskip\pFqmuskip
\newcommand*\pFq[6][8]{%
	\begingroup 
	\pFqmuskip=#1mu\relax
	\mathchardef\normalcomma=\mathcode`,
	\mathcode`\,=\string"8000
	\begingroup\lccode`\~=`\,
	\lowercase{\endgroup\let~}\pFqcomma
	{}_{#2}F_{#3}{\left[\genfrac..{0pt}{}{#4}{#5};#6\right]}%
	\endgroup
}
\newcommand{\pFqcomma}{{\normalcomma}\mskip\pFqmuskip}

\ExplSyntaxOn
\NewDocumentCommand{\MeijerG}{smmmm}
{
	\IfBooleanTF{#1}
	{
		\vic_meijerg:nnnnnn { #2 } { #3 } { #4 } { #5 } { small } { }
	}
	{
		\vic_meijerg:nnnnnn { #2 } { #3 } { #4 } { #5 } { } { \; }
	}
}

\seq_new:N \l__vic_meijerg_args_in_seq
\seq_new:N \l__vic_meijerg_args_out_seq

\cs_new_protected:Nn \vic_meijerg:nnnnnn
{
	\seq_set_split:Nnn \l__vic_meijerg_args_in_seq { | } { #3 }
	\seq_clear:N \l__vic_meijerg_args_out_seq  
	\seq_map_inline:Nn \l__vic_meijerg_args_in_seq
	{
		\seq_put_right:Nn \l__vic_meijerg_args_out_seq
		{
			\begin{#5matrix} ##1 \end{#5matrix}
		}
	}
	G\sp{#1}\sb{#2}
	\left(
	\seq_use:Nn \l__vic_meijerg_args_out_seq { #6\middle|#6 }
	#6\middle|#6
	#4
	\right)
}
\ExplSyntaxOff

\makeatother
\begin{document}
\title{Wannier Excitons Confined in Hexagonal Boron Nitride Triangular Quantum Dots}
\author{M. F. C. Martins Quintela}
\email{mfcmquintela@gmail.com}
\affiliation{Department of Physics and Physics Center of Minho and Porto Universities (CF--UM--UP), Campus of Gualtar, 4710-057, Braga, Portugal}
\affiliation{International Iberian Nanotechnology Laboratory (INL), Av. Mestre Jos{\'e} Veiga, 4715-330, Braga, Portugal}
\author{N. M. R. Peres}
\affiliation{Department of Physics and Physics Center of Minho and Porto Universities (CF--UM--UP), Campus of Gualtar, 4710-057, Braga, Portugal}
\affiliation{International Iberian Nanotechnology Laboratory (INL), Av. Mestre Jos{\'e} Veiga, 4715-330, Braga, Portugal}
\begin{abstract}
	With the ever-growing interest in quantum computing, understanding the behaviour of excitons in monolayer quantum dots has become a topic of great relevance. In this paper, we consider a Wannier exciton confined in a triangular quantum dot of hexagonal Boron Nitride. We begin by outlining the adequate basis functions to describe a particle in a triangular enclosure, analyzing their degeneracy and symmetries. Afterwards, we discuss the excitonic hamiltonian inside the quantum dot and study the influence of the quantum dot dimensions on the excitonic states.
\end{abstract}
\maketitle

\section{Introduction}

Recently, the study of monolayer quantum dots and their optical properties has gained increased traction due do their interest in quantum computing\cite{Krenner_2005,Maragkou_2015}. The control and properties of single electrons and electron--hole pairs in qubits, the strong spin--valley coupling\cite{PhysRevX.4.011034,Mak2018}, the possibility of electrically confining and manipulation of charge carriers\cite{Wang2018}, and the control of valley splitting and polarization of excitons\cite{Qu_2019}, together with the possibility of optically controlling two--qubit operations in a single quantum dot \cite{Wu2016}, makes understanding excitons confined in quantum dots extremely relevant.

Further advances in understanding the triangular growth patterns of monolayer transition metal dichalcogenide (TMD) monolayers\cite{Zhu2015} led to improvements in the mechanical deposition of triangular two--dimensional TMD crystals\cite{GovindRajan2016}. Much more recently, methods based on the combination of periodic laser patterns, anisotropic thermal etching and endoepitaxial growth enabled the realization of monolayer nanostructures with atomically sharp interfaces\cite{Liu2022}. The stability of triangular quantum dots has also been considered for both regular TMDs \cite{10.1021/acsanm.8b00634} and Janus materials\cite{Paez-Ornelas2021}, where the chalcogen layers above and below the transition metal layer are composed of different materials\cite{Lu2017,Zhang2017,10.1063/1.5135306}. 

Optical, electronic and magnetic properties of transition metal dichalcogenide quantum dots are of special interest for quantum computing. Various recent works have focused on the study of these properties via different methods, namely tight--binding\cite{PhysRevB.93.161404,PhysRevB.94.245429,Avalos_Ovando_2018}, density functional theory\cite{PhysRevX.4.011034,Tiutiunnyk2022} and continuum models\cite{PhysRevB.81.201402,PhysRevB.95.045409,Mittelstadt2022}. Phase transitions to the metallic phase are also especially relevant, allowing for different charge excitation and decay pathways\cite{Kim2021}. Recent experimental works have also shown that chemical treatment is a effective method for enhancing and modulating exciton emission in triangular TMD quantum dots \cite{Dhakal2017}. Triangulenes, triangular graphene flakes with a lateral dimension of $ n $ benzene rings, have also been a recent point of focus due to their electronic and spintronic properties\cite{10.1126/sciadv.aav7717,10.1002/anie.202108301,10.48550/arxiv.2206.14907}.

\begin{figure}
	\includegraphics[scale=0.8]{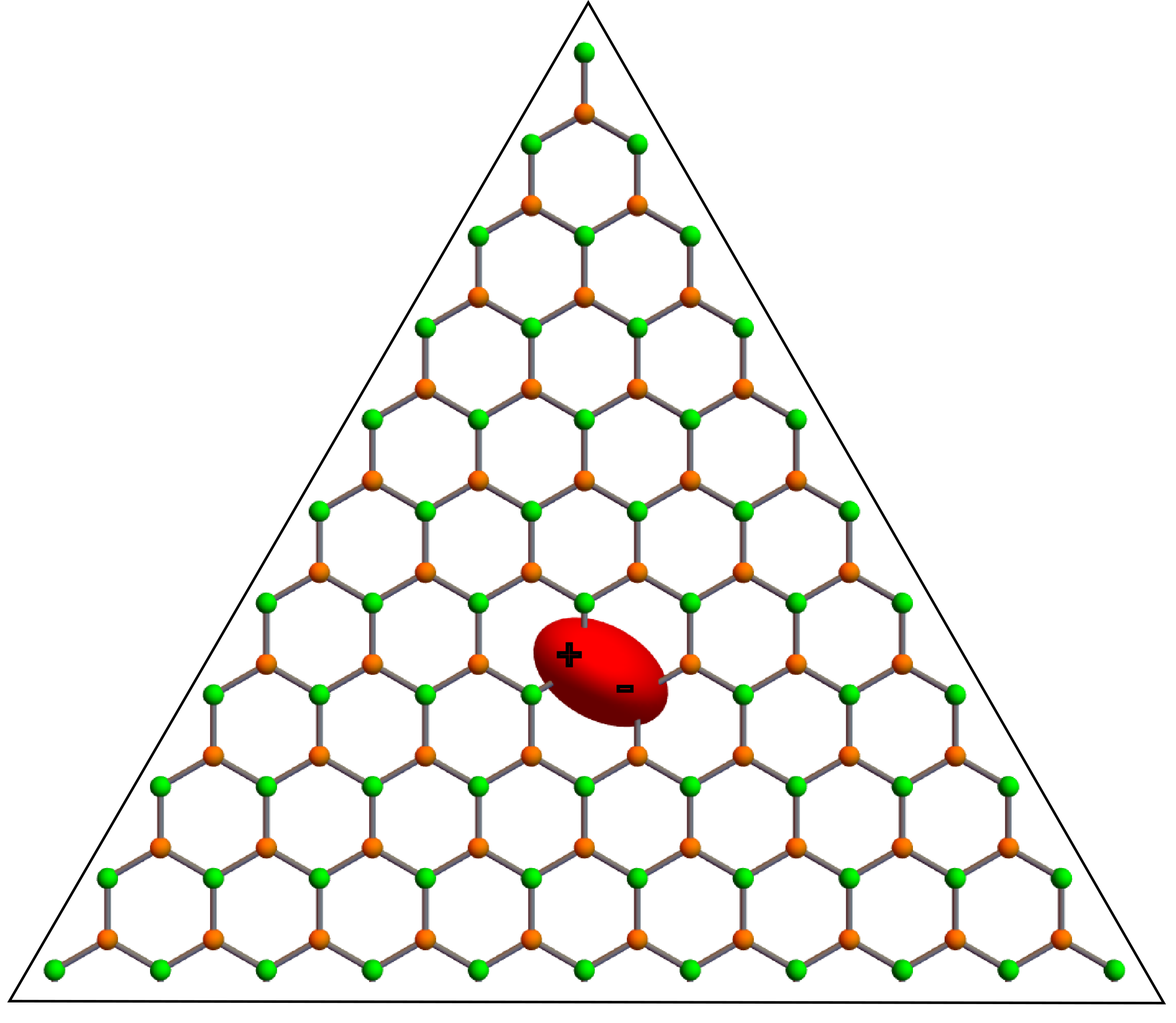}
	\caption{Schematic view of an exciton confined in a hBN triangular quantum dot with zigzag boundaries. \label{fig:lattice}}
\end{figure}

In this paper, we consider the simple system of Wannier excitons in a hexagonal Boron Nitride triangular quantum dot (Fig. \ref{fig:lattice}). Using a finite basis method, we obtain analytical approximations to the excitonic wave functions before considering the effects of varying the size of the confinement region. This paper is structured as follows. In Sec. \ref{sec:triangl_particle}, we begin by describing the basis of functions we will consider. This basis is created from the exact solutions for the problem of a particle in an infinite triangular well. We then perform a simple symmetry analysis of this basis such that we can more easily compute and understand the Hamiltonian for the excitonic system. In Sec. \ref{sec:wannier_exciton_dot}, we turn our attention to the Wannier equation in a triangular quantum dot, discussing the electrostatic interaction between the electron and hole. Focusing our attention on hexagonal Boron Nitride (hBN), we compute the effective masses from the band structure before studying the convergence of the method as the basis size grows. Finally, in Sec. \ref{sec:dot_size}, we vary the size of the triangular quantum dot to understand the effects of confinement on each individual excitonic state.

\section{Triangular Quantum Dots\label{sec:triangl_particle}}

We will first begin by considering a particle confined in an equilateral triangle with hard boundaries. The eigenvalue problem of the Schrödinger equation for an equilateral triangle with hard boundaries was solved by Lamé \cite{Lame1852} in 1852. As in the well known case of a rectangle, this solution is a finite sum of plane waves, built to satisfy the boundary conditions. It has been proven that the only polygons that share this feature are the square, the rectangle, the equilateral triangle, the half-square (right isosceles triangle) and the half equilateral triangle ($ 30^\circ - 60^\circ $ triangle) \cite{ISI:A1991GC36900022,ISI:A1993LQ34500002}. 

\subsection{Analytical Solution of Triangular Well}

The solution to the particle in the triangular well has been studied through various different lenses, resulting in somewhat different expressions that all turn out to be equivalent\cite{RICHENS1981495,Krishnamurthy_1982,Turner_1984,doi:10.1063/1.526701,doi:10.1137/S003614450238720,Gaddah_2013}. We will focus our studies on the solution provided by Gaddah\cite{Gaddah_2013}, developed for an equilateral triangle of side $ L $ with vertices $ A,\;B,\;C $ located at $ \left( 0,0\right)  $, $ \left( L,0\right)  $ and $ \left( L/2, \sqrt{3}L/2\right)  $, respectively. The Schrödinger equation for a particle of mass $ \mu $ confined in the infinite triangular well defined by the vertices $ A,\;B,\;C $ is written as
\begin{equation}\label{eq:schro}
	-\frac{\hbar^{2}}{2 \mu}\left(\frac{\partial^{2}}{\partial x^{2}}+\frac{\partial^{2}}{\partial y^{2}}\right) \Psi(x, y)=E \Psi(x, y),
\end{equation}
and subjected to the Dirichlet condition that $ \Psi(x, y) $ vanishes on the perimeter of the triangle. 

Applying a procedure based on group theory\cite{Gaddah_2013}, the eigenfunctions are split into two different functional classes
\begin{equation}
	\Psi_{n, m}^{(j)}(x, y)= 
	\begin{cases}
		N_{1} \operatorname{Re}\left\{W\left(z, z^*\right)\right\}, & j=1 \\ 
		N_{2} \operatorname{Im}\left\{W\left(z, z^*\right)\right\}, & j=2,
	\end{cases}
\end{equation}
where $ z=x+iy $ and $ N_j $ are normalization constants. The function $ W\left(z, z^*\right) $ can be expressed as the determinant of a $ 3\times3 $ matrix as
\begin{widetext}
\begin{equation}
	W(x+i y, x-i y)=\left|
		\begin{array}{ccc}
			1 & 1 & 1 \\
			\mathrm{e}^{\mathrm{i} \beta n(x+\sqrt{3} y)} & \mathrm{e}^{\mathrm{i} \beta n(x-\sqrt{3} y)} & \mathrm{e}^{-2 \mathrm{i} \beta n x} \\
			\mathrm{e}^{-\mathrm{i} \beta m(x+\sqrt{3} y)} & \mathrm{e}^{-\mathrm{i} \beta m(x-\sqrt{3} y)} & \mathrm{e}^{2 \mathrm{i} \beta m x}
		\end{array}\right|,
\end{equation} 
where $ \beta=2\pi/3L $. Expanding the two different functional classes, the eigenstates are given by 
\begin{align}
	\Psi_{n, m}^{(1)}(x, y)&=2 N_{1} \left[ \sin \left( \beta \left( m-n \right) x \right) \sin \left(\sqrt{3} \beta\left(n+m\right) y\right)+\sin \left(\beta\left(n+2 m\right) x\right) \sin \left(\sqrt{3} \beta n y\right)-\right. \nonumber\\
	&\quad \left. -\sin \left(\beta \left(2 n+m\right) x\right) \sin \left(\sqrt{3} \beta m y\right)\right]\label{eq:basis_1} \\
	\Psi_{n, m}^{(2)}(x, y)&=2 N_{2}\left[\cos \left(\beta\left(m-n \right) x\right) \sin \left(\sqrt{3} \beta\left(n +m \right) y\right)-\cos \left(\beta\left(n+2 m \right) x\right) \sin \left(\sqrt{3} \beta n y\right)-\right. \nonumber \\
	&\quad \left. -\cos \left(\beta \left( 2 n+m \right) x\right) \sin \left(\sqrt{3} \beta m y\right)\right].\label{eq:basis_2}
\end{align}
\end{widetext}
The quantum numbers $ n,\;m $ obey the restriction 
\begin{equation}
	m\geq n>0,
\end{equation}
with the values $ m=n $ only allowed for the functions $ \Psi_{n, m}^{(2)}(x, y) $ as they lead to the trivial zero solution for $ \Psi_{n, m}^{(1)}(x, y) $.
The orthogonality of these eigenfunctions is ensure by both Rellich’s theorem \cite{MacCluer1994-hl}, for eigenfunctions corresponding to distinct eigenvalues, and by direct integration of the different eigenfunctions when the eigenvalues are degenerate. 

The energy eigenvalues are given by 
\begin{equation}
	E_{n, m}=\frac{8 \pi^{2} \hbar^{2}}{9 \mu L^{2}}\left(n^{2}+n m+m^{2}\right).
\end{equation}
Since both eigenfunctions $ \Psi_{n, m}^{(1)},\; \Psi_{n, m}^{(2)} $ correspond to the same energy eigenvalue $ E_{n, m} $, then it follows that all eigenvalues corresponding to $ n\neq m $ are (at least), $ 2\times $ degenerate. Higher order degeneracy can also occur accidentally, as well as in cases where $ n=m $ (\emph{e.g.}, the state $ (n,m)=(7,7) $ is degenerate with the two states corresponding to $ (n,m)=(2,11) $).

\subsection{Symmetry Analysis of Truncated Basis\label{sec:symmetry_analysis}}
In order to truncate the eigenfunction basis, we will need to sort them in some way. To do this in a reasonable manner, we will sort the eigenfunctions by increasing energy eigenvalue. This can be easily done, as the set of all acceptable values of $ n,\;m $ can be expressed as 
\begin{equation}
	m=n+s,\qquad  n=1,2,3,...,
\end{equation}
with $ s\in \left\{1,2,3,...\right\} $ for $ \Psi_{n, m}^{(1)}(x, y) $ and $ s\in \left\{0,1,2,3,...\right\} $ for $ \Psi_{n, m}^{(2)}(x, y) $, allowing us to quickly run through many basis functions in an algorithmic way. 

From a group symmetry perspective, each $ \Psi_{n, m}^{(1/2)}(x, y) $ will have certain symmetries that will allow us to diagonalize the Hamiltonian in blocks, greatly speeding up the computation time and reducing the number of trivially--zero matrix elements that are computed to zero. These eigenstates are, however, not written in the usual irreducible representations of the $ C_{3v} $ point group, so a more careful analysis is required. By careful inspection of the eigenfunctions of the infinite triangular well, four different and orthogonal groupings of functions can be observed. For this discussion, we define non--negative integer $ k $ to better distinguish the functions. Firstly, the fully symmetric states are all those written as
\begin{equation}
	\Psi_{n, n+3k}^{(2)}(x, y).
\end{equation}
Secondly, the remaining $ \Psi_{n, n+s}^{(2)}(x, y) $ (\emph{i.e., } those where $ s\mod 3 = 1 $ or $ 2 $) all belong to a separate grouping in which all functions are even under reflection by the axis that crosses both the origin and the center of the triangle $ \left(\frac{L}{2}, \frac{L}{2\sqrt{3}}\right) $.
Thirdly, all eigenstates written as 
\begin{equation}
	\Psi_{n, n+3k}^{(1)}(x, y)
\end{equation}
are even under rotations of $ \pm 2\pi /3 $ along the vertical axis that passes through the geometric center of the triangle.
Fourthly and finally, the remaining $ \Psi_{n, n+s}^{(1)}(x, y) $ (\emph{i.e., } those where $ s\mod 3 = 1 $ or $ 2 $) are all odd under reflections by the axis that crosses both the origin and the center of the triangle. This separation matches what would be expected from the energy degeneracy of the states $ \Psi_{n, m}^{(1/2)}(x, y) $, as these belong to orthogonal sets of functions.

With this block diagonalization of the Hamiltonian, we can now move to studying to convergence of this method. To do this, however, we will first introduce a notation for the states mentioned previously as to make the discussion easier to follow. The fully symmetric states will be labeled as 
\[
\phi_{\mathrm{Id}}(x, y), 
\]
the states even/odd under reflection by the axis that crosses both the origin and the center of the triangle are labeled as
\[
\phi_{\pm \sigma}(x, y),
\]
respectively, and the states even under rotations of $ \pm 2\pi /3 $ along the vertical axis that passes through the geometric center of the triangle are labeled as 
\[
\phi_{C_3}(x, y).
\]
The quantum numbers $ n,s $ have been discarded in this section for two specific reasons. Firstly, to improve readability and clarity when discussing the states and the convergence of the method; secondly, because the functions in each orthogonal grouping will be sorted by their energies for the implementation of the method, effectively removing the meaning of each individual quantum number.

\section{Wannier Equation in a Quantum Dot\label{sec:wannier_exciton_dot}}

Following from the Schr{\"o}dinger in a triangular well in Eq. (\ref{eq:schro}), we will now consider the Wannier equation inside a hexagonal Boron Nitride triangular dot, given by 
\begin{equation}\label{eq:wannier}
	-\frac{\hbar^{2}\nabla^{2}}{2 \mu_{\mathrm{exc}}}\psi_{\nu}\left(\mathbf{r}\right)+ V\left(\mathbf{r}\right) \psi_{\nu}\left(\mathbf{r}\right)=E_{\mathrm{bind},\nu} \psi_{\nu}\left(\mathbf{r}\right),
\end{equation}
where $ \nabla^{2} $ is the Laplacian operator, $ \mu_{\mathrm{exc}} $ is the exciton reduced mass, $ \psi_{\nu}\left(\mathbf{r}\right) $ the wave function for the excitonic state $ \nu $, $ E_{\mathrm{bind},\nu}=E_{\nu}-E_{\mathrm{gap}} $ the binding energy associated with the excitonic state $ \psi_{\nu}\left(\mathbf{r}\right) $, with $ E_{\nu}$ the energy of the state and $E_{\mathrm{gap}} $ the bandgap of the system, and $ V\left(\mathbf{r}\right) $ an electrostatic potential coupling the electron and the hole pair. 

To obtain the excitonic binding energies and wave functions, we expand $ \psi_{\nu}\left(\mathbf{r}\right) $ in terms of the solutions of the infinite well defined in Eqs. (\ref{eq:basis_1}--\ref{eq:basis_2}). As the electrostatic potential coupling the electron and hole pair only depends on the distance to the center of the triangular dot, the excitonic wave functions will have the same symmetry as the states $ \phi_{\mathrm{Id}},\phi_{\pm \sigma},\phi_{\mathrm{C_{3}}} $. This means that Eq. (\ref{eq:wannier}) can also be diagonalized by blocks for the basis size chosen and, therefore, the wave functions can be explicitly separate by their symmetries and written as  
\begin{align}\label{eq:expansion}
	\psi_{\nu,\mathrm{S}}\left(\mathbf{r}\right)&=\sum_{n=0}^{N}c_{n,\mathrm{S}}\,\phi_{n,\mathrm{S}}\left(\mathbf{r}\right),
\end{align}
where $ N $ is the basis size chosen, $ \left\{c_n\right\} $ are a set of numerical coefficients yet to be determined, and $ \mathrm{S}=\left\{\mathrm{Id},\mathrm{+\sigma},\mathrm{-\sigma},\mathrm{C_{3}}\right\} $ represents the symmetry of the excitonic state following the analysis from Sec. \ref{sec:symmetry_analysis}.

To compute the reduced exciton mass $ \mu_{\mathrm{exc}} $, we start from the low--energy Hamiltonian, given by 
\begin{equation}\label{eq:hamilt_hbn}
	\mathcal{H}\left(\mathbf{k}\right)=\frac{E_{g}}{2} \sigma_{z}+\hbar v_{F} \left( \tau k_{x} \sigma_{x}+ k_{y} \sigma_{y}\right),
\end{equation}
where $ 2E_g=7.8\,\mathrm{eV} $ is the bandgap, $ \sigma_{i} $ are the Pauli matrices, $ \tau=\pm 1 $ is the valley index and $ v_F $ is the Fermi velocity. The band structure near the Dirac point is given by 
\begin{equation}\label{eq:band_structure}
	E_{\pm}\left(\mathbf{k}\right)=\pm \sqrt{\frac{E_g ^{2}}{4}+\hbar^{2} v_{F}^{2} k^{2}},
\end{equation}
near $ k=0 $, where $ \pm $ denotes the conduction/valence bands. The reduced mass of excitons in this material can be obtained from the coefficient of the quadratic term of the series expansion of the band structure present in Eq. (\ref{eq:band_structure}) near $ k=0 $ \cite{Quintela_tutorial}. Performing this expansion, the relevant coefficient is given by
\begin{equation}\label{eq:reduced_mass}
	\begin{aligned}
		\frac{\hbar^{2}}{m_{\pm}}&=\left. \frac{\partial^{2}E_{\pm}\left(\mathbf{k}\right)}{\partial k^{2}}\right|_{k=0}\\
		&= \pm \frac{2 \hbar^{2} v_{F}^{2}}{E_g}.
	\end{aligned}
\end{equation}
The reduced exciton mass is then obtained as $ \mu_{\mathrm{exc}}^{-1}=m_{+}^{-1}-m_{-}^{-1}$, meaning that
\begin{equation}\label{eq:exciton_mass}
	\mu_{\mathrm{exc}}=\frac{E_g}{4 v_{F}^{2}}.
\end{equation}

\subsection{Electrostatic Potential in Layered Materials}

For two dimensional layered materials, the electrostatic coupling between the electron and the hole of the exciton is well--modeled\cite{Trolle2017} by the Rytova--Keldysh potential\cite{rytova1967,keldysh1979coulomb}, given by \begin{equation}
	V\left(\mathbf{r}\right)=-\frac{\hbar c \alpha \pi}{2 r_{0}}\left[H_{0}\left(\epsilon_{r}\frac{r}{r_{0}}\right)-Y_{0}\left(\epsilon_{r}\frac{r}{r_{0}}\right)\right]
	\label{eq:Keldysh_potential},
\end{equation}
where $ r $ is the in--plane position, $\epsilon_{r}$ is the mean dielectric constant of the medium above/below the layered material, $ c $ is the speed of light, $ \alpha $ is the fine--structure constant, and $ H_0 (x) $ and $ Y_0 (x) $ are the Struve function and Bessel function of the second kind, respectively. The parameter $r_{0}$ corresponds to an in--plane screening length related to the 2D polarizability of the material, usually obtained via DFT\cite{acs.nanolett.9b02982}.

\begin{figure}
	\includegraphics[scale=0.95]{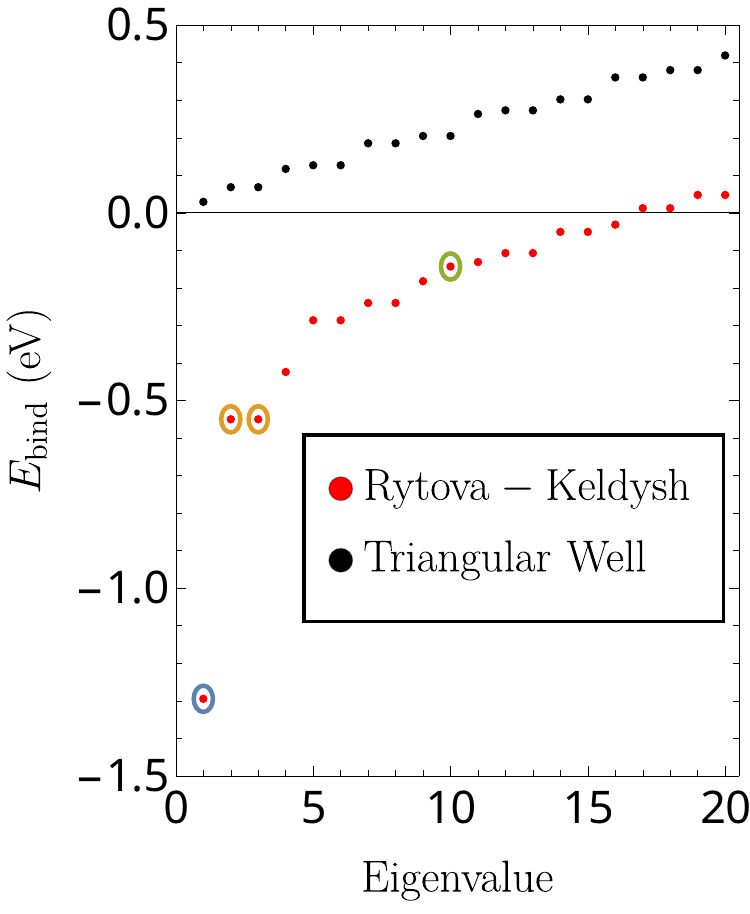}
	\caption{Spectrum of a particle confined in an infinite triangular well (black dots) and of a Wannier exciton confined to a triangular quantum dot (red dots) for a basis size of $ N=80 $ and a side length $ L=200\,\text{\AA} $. Colored circles mark the lowest energy states from $ \phi_{\mathrm{Id}} $ (blue), $ \phi_{\pm \sigma} $ (orange) and $ \phi_{C_3} $ (green).\label{fig:method_eigenval}}
\end{figure}

Considering again our triangular dot, the hard wall confinement leads to a Dirichlet boundary condition on $ \psi_{\nu}\left(\mathbf{r}\right)=0 $, meaning that the basis of functions defined in Eqs. (\ref{eq:basis_1}-\ref{eq:basis_2}) is an appropriate choice for a finite basis approach. Additionally, we shift the Rytova--Keldysh potential as to be centered at the geometric center of our triangular dot, meaning that the electrostatic potential inside it now reads 
\begin{equation}
	V\left(\mathbf{r}\right)=-\frac{\hbar c \alpha \pi}{2 r_{0}}\left[H_{0}\left(\epsilon_{r}\frac{r^{\prime}}{r_{0}}\right)-Y_{0}\left(\epsilon_{r}\frac{r^{\prime}}{r_{0}}\right)\right]
	\label{eq:Keldysh_potential_centered},
\end{equation}
with $ r^{\prime}=\left|r-r_{c}\right|=\sqrt{\left(x-x_{0}\right)^{2}+\left(y-y_{0}\right)^{2}} $ the distance between a point $ r=(x,y) $ and the center of the triangle $ r_c = \left(x_0 , y_0\right) $. For freely suspended hBN, the screening length parameter in the Rytova--Keldysh potential is $ r_0 = 10\,\text{\AA}$ \cite{henriques_optical_2020} and $ \epsilon_{r}=1 $.

\subsection{Convergence Analysis}
\begin{figure}
	\includegraphics[scale=0.95]{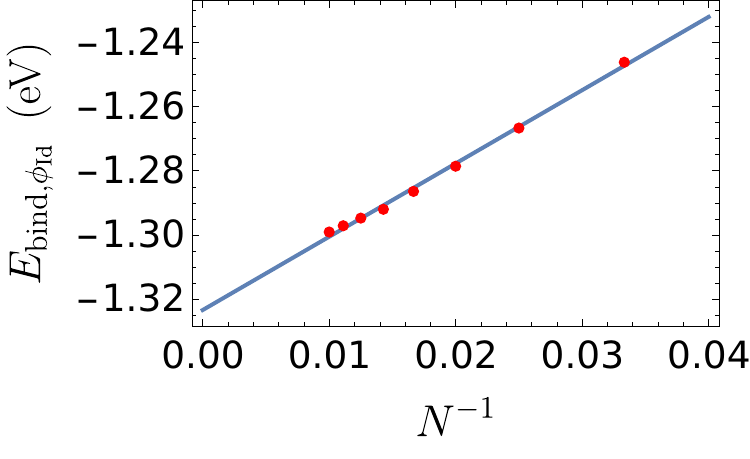}
	\caption{Scaling of the binding energy of the lowest $ \phi_{\mathrm{Id}} $ state as a function of $1/N$, with $N\in[30,100]$, for $ L=200\,\text{\AA} $. The interception of the fitting straight--line with energy axis gives a ground state energy $E_{\phi_{\mathrm{Id}}}=-1.323\,\mathrm{eV}$. \label{fig:method_convergence}}
\end{figure}

\begin{figure*}
	\includegraphics{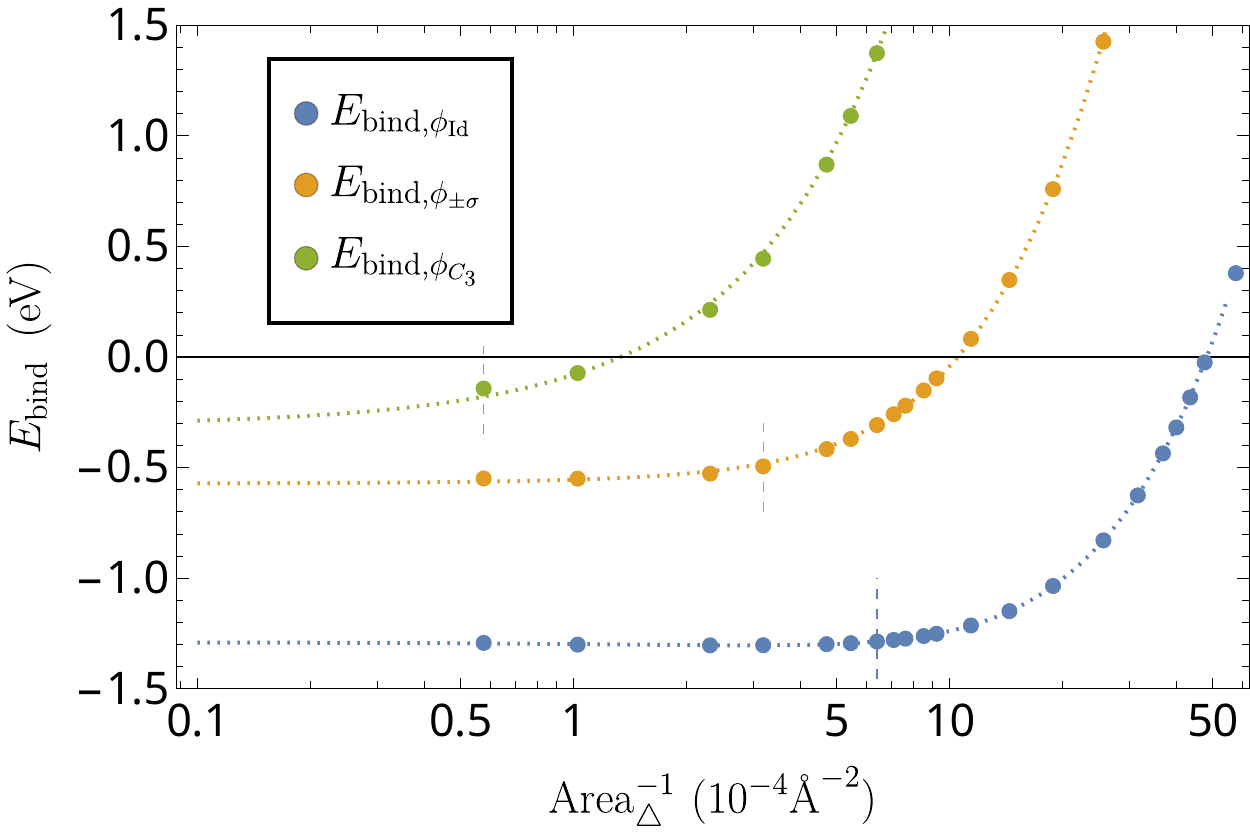}
	\caption{Scaling of the binding energy of the lowest $ \phi_{\mathrm{Id}} $, $ \phi_{\pm \sigma} $ and $ \phi_{C_3} $ states as a function of the inverse of the area of the triangular dot for a side length range $ L\in[20,200]\,\text{\AA} $. Vertical ticks mark the area for which the specific state begins to feel confinement.\label{fig:method_Lchange}}
\end{figure*}

Having defined both the electrostatic potential and the reduced exciton mass in the system, we can proceed to compute the four distinct Hamiltonian blocks for each of the orthogonal representations for a specific basis size $ N $ and a specific side length $ L $. Doing this, we compare the full low energy spectrum of the exciton confined in a triangular dot to that of the simple particle inside a triangular box in Fig. (\ref{fig:method_eigenval}).

Focusing on the lowest energy state, belonging, as expected, to the $ \phi_{\mathrm{Id}} $ set of functions, we study the evolution of its energy eigenvalue as the basis size increases. In Fig. (\ref{fig:method_convergence}), we plot this evolution as a function of the inverse of the basis size for a triangle side length of $ L=200\,\text{\AA} $. We obtain the expected limiting value for the complete basis by the intersection of the fitting line with the axis as $E_{\phi_{\mathrm{Id}}}=-1.323\,\mathrm{eV}$.

\section{Quantum Dot Size Effects\label{sec:dot_size}}

\begin{figure*}
	\centering
	\includegraphics[scale=0.9]{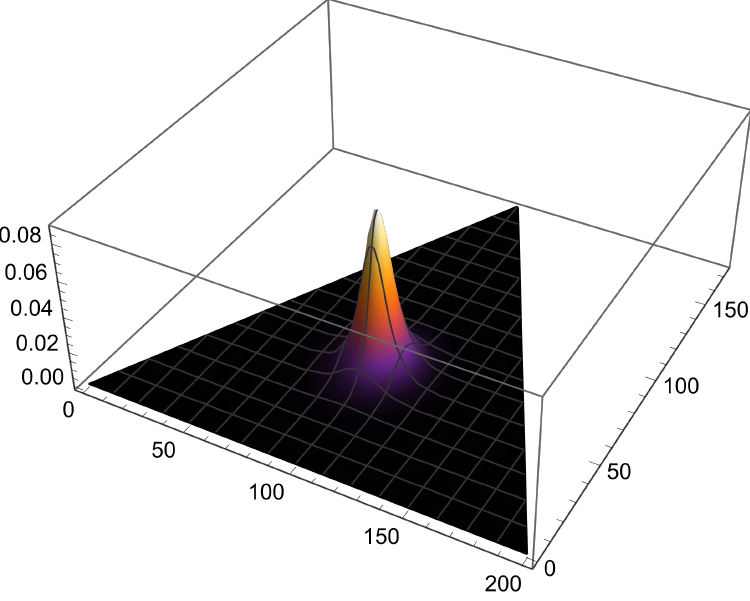}
	\qquad\qquad\includegraphics[scale=0.9]{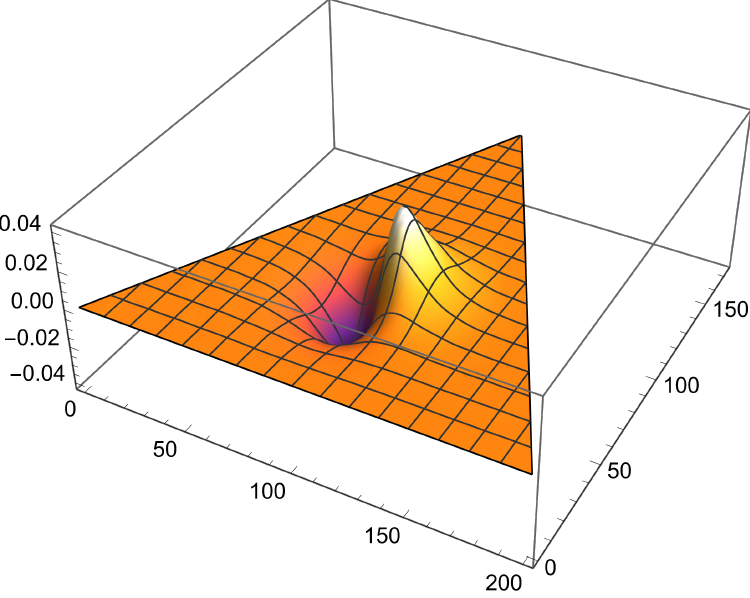}
	
	\includegraphics[scale=0.9]{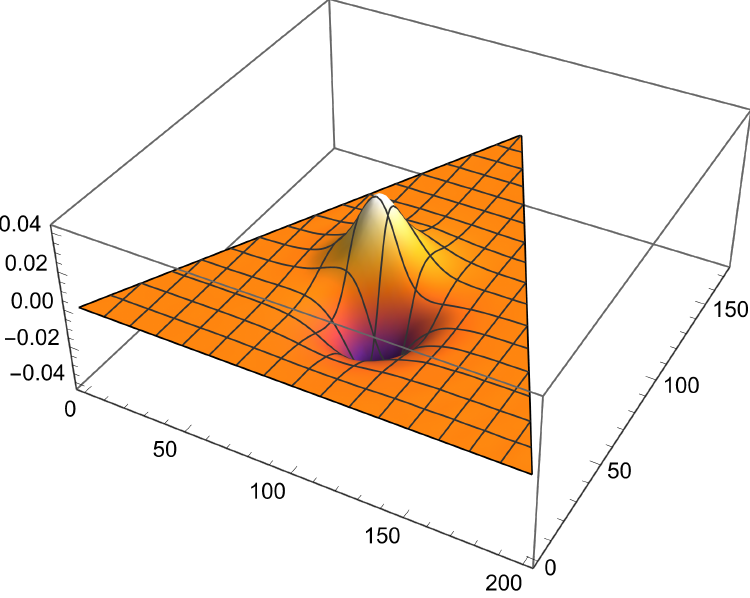}
	\qquad\qquad\includegraphics[scale=0.9]{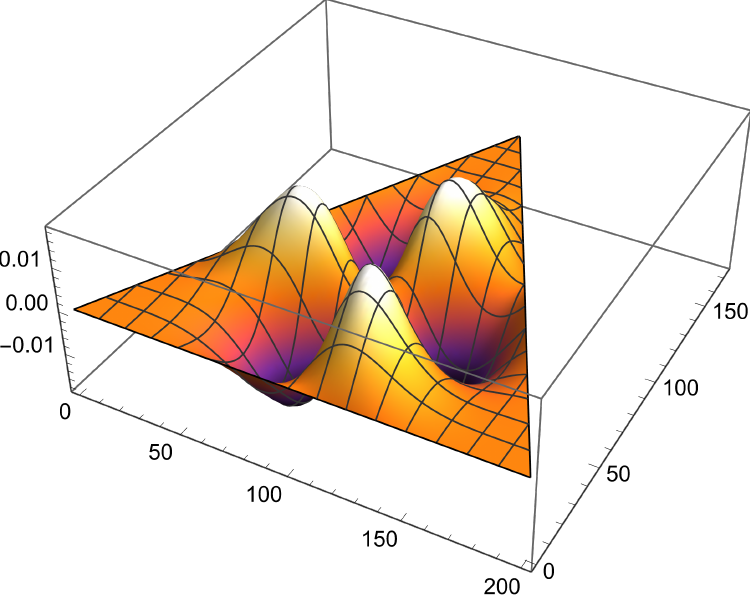}
	
	\caption{3D plot of the lowest energy excitonic wave functions belonging to $ \phi_{\mathrm{Id}} $ (top-left), $ \phi_{\mathrm{+\sigma}} $ (top-right), $ \phi_{\mathrm{-\sigma}} $ (bottom-left), and $ \phi_{\mathrm{C_{3}}} $ (bottom-right) confined in a hBN triangular quantum dot of side $ L=200\text{\AA} $ considering a basis size $ N=80 $.\label{fig:3d_plots}}
\end{figure*}

To finalize this paper, we will now discuss the effect on the excitonic states of varying the size of the quantum dot. This is done by setting the basis size to a fixed value, in this case $ N=80 $ as to greatly speed up the computation versus $ N=100 $, and varying the side length of the triangular dot. To study this dependence, we plot the energy of the excitonic ground state as a function of the inverse of the area $ \left(=\frac{\sqrt{3}L^{2}}{4}\right) $ of the triangular dot in Fig. (\ref{fig:method_Lchange}). We choose this scaling as it is the same relation as that between the energy of a particle in the textbook problem of hard wall confinement in a polygonal box with hard wall confinement and the dimensions of said box (\emph{i.e.,} the energy of a state will be inversely proportional to the area of the box). To aid visualization, the states in question have also been marked in Fig. (\ref{fig:method_eigenval}) with their respective colors. The wave functions of these states are also plotted in Fig. (\ref{fig:3d_plots}).

As the confinement effects of the triangular dot start dominating the excitonic states as the dot size shrinks, we expect a linear trend in the ground state energy as we reduced the size of the triangle. As the dot size grows, however, we expect the energy to settle to an almost constant value, as for a sufficiently large quantum dot no effects from the boundaries are felt. This is displayed in the vertical tick marks present in Fig. (\ref{fig:method_Lchange}) marking the points at which the confinement effects become noticeable. For the excitonic ground state and first excited state, this happens at $ L\approx60\,\text{\AA} $ and $ L\approx85\,\text{\AA} $, respectively. For the first state belonging to the $ \phi_{C_3} $ block, this occurs at a side length superior to $ L\approx200\,\text{\AA} $. This is as expected, as the lowest energy state from the $ \phi_{C_3} $ block is the sixth excited state in the total Hamiltonian, as is clear from Fig. (\ref{fig:method_eigenval}).

A relevant comparison for the various side lengths is against the distance between edge atoms in the triangular dot. Considering a zigzag boundary, where the edges of the dot are all composed of the same atomic site (Boron or Nitrogen), the distance between two atoms in the edge is given by 
\begin{align}
	d&=2a\cos30\degree=4.347\,\text{\AA},
\end{align}
where $ a=2.510\,\text{\AA} $ is the lattice parameter of hBN. As such, an edge with length $ L $ will have $ L/d $ atoms on the edge. 

Finally, we will now consider both the ground state and the two times degenerate first excited state. As these three ($ 1+2 $) states are analogous to the $ 1s $ and $ 2p_{\pm} $ states in the 2D Hydrogen atom, we expect them to correspond to the dominant resonance in the excitonic polarizability\cite{frohlich_observation_1985,berghauser_optical_2016,berghauser_mapping_2018,acs.nanolett.9b02982}, easily accessible via pump--probe experiments\cite{poellmann_resonant_2015,Merkl2019}. The ground state begins to feel the effects of the triangular confinement at $ L\approx60\,\text{\AA} $, which corresponds to a hBN triangular dot whose sides consist of roughly $ 14 $ atoms. The first excited state only feels the confinement effects at $ L\approx85\,\text{\AA} $, corresponding to an edge of roughly $ 20 $ atoms. This creates a range of side lengths in which the excitonic ground state already considers the system infinite, but the first excited states are still feeling the effects of the triangular confinement. 

\section{Conclusions}

In this paper, we have studied the properties of excitons confined to triangular quantum dots. We began by discussing an appropriate basis of functions to obtain an approximate solution to the Wannier equations. These functions follow from the solution of the Schrödinger equation for a particle confined to a triangular well, where analytical closed form solutions are well known. Afterwards, we briefly analyzed the symmetry of these wave functions so that we could more efficiently compute the energy expectation values and, therefore, reduce the computational time. This symmetry analysis provides valuable information about the degeneracy of each excitonic state, as well as understanding which dipolar transitions one can expect to be allowed. 

Afterwards, we considered Wannier excitons confined in a hexagonal Boron Nitride quantum dot, explicitly writing the electrostatic coupling between electron and hole as well as their respective reduced masses. The convergence of the method with increasing basis size was also discussed.

Finally, we considered the effects of decreasing the size of the quantum dot on the excitonic states. As expected, the more energetic states are more sensible to confinement as they are more spread out over the dot. We focused our attention on the two lowest energy states, analogous to the Hydrogen $ 1s $ and $ 2p $ states, as transitions between these are expected to dominate the excitonic linear polarizability. 

\section*{Acknowledgements}

M.F.C.M.Q. acknowledges the International Nanotechnology Laboratory (INL) and the Portuguese Foundation for Science and Technology (FCT) for the Quantum Portugal Initiative (QPI) grant SFRH/BD/151114/2021. 
N.M.R.P. acknowledges support by the Portuguese Foundation for Science and Technology (FCT) in the framework of the Strategic Funding UIDB/04650/2020, COMPETE 2020, PORTUGAL 2020, FEDER, and  FCT through projects PTDC/FIS-MAC/2045/2021, EXPL/FIS-MAC/0953/ 2021, and from the European Commission through the project Graphene Driven Revolutions in ICT and Beyond (Ref. No. 881603, CORE 3).

\bibliographystyle{unsrt}
\bibliography{biblio_TMD_QD}

\end{document}